\documentclass[a4paper,11pt]{article}
\usepackage{pos}
\usepackage{aasmacros}
\usepackage{url}
\usepackage[utf8]{inputenc}
\usepackage{soul}

\title{The relevance of the cosmic microwave background for cosmology}

\author*[a,b]{Pavel Kroupa}
\author[c,d]{Eda Gjergo}
\author[e]{Nikolaos Samaras}

\affiliation[a]{Helmholtz Institute for Radiation and Nuclear Physics, University of Bonn\\
  Nussallee 14-16, 53115 Bonn, Germany}
\affiliation[b]{Astronomical Institute, Charles University\\
V Holesovickach 2, 18000 Prague, Czech Republic, CZ-18000 Praha, Czech Republic}
\affiliation[c]{School of Astronomy and Space Science, Nanjing University, Nanjing 210023, People’s Republic of China}
\affiliation{[d]Key Laboratory of Modern Astronomy and Astrophysics (Nanjing University), Ministry of Education, Nanjing 210093, People’s Republic of China}
\affiliation[e]{Department of Physics, University of Ottawa, Ottawa, Canada K1N 6N5}

\emailAdd{pkroupa@uni-bonn.de}
\emailAdd{eda.gjergo@nju.edu.cn}
\emailAdd{nsamaras@uottawa.ca}

\abstract{Within the Hot Big Bang picture, as usually interpreted in the standard $\Lambda$CDM framework and its hierarchical scenario for elliptical galaxy assembly, the cosmic microwave background (CMB) is associated with photon decoupling near $z \approx 1100$, and the Planck angular power spectra are taken to constrain primordial $10^{-5}$-level fluctuations that seeded later structure formation. We show that, in addition to the $\Lambda$CDM model, two cosmological models based on Milgromian gravitation are also consistent with the Planck CMB power spectra, {\it such that the spectra do not uniquely favor dark matter}. We then argue that the published CMB spectra are not direct observables, but reconstructed quantities obtained after foreground subtraction and corrections for weak lensing, photon-electron scattering, and gravitational energy redshifts along the line of sight. Because these steps depend on the assumed growth history of structure, {\it the resulting spectra depend on the cosmological model adopted in the data reduction}.  
We also note that the reported correlation between the hemispherical CMB asymmetry and the excess on-sky distribution of elliptical galaxies may point to matter inhomogeneities spanning large portions of the observable universe.
Motivated by observational tensions in the hierarchical assembly of elliptical galaxies, we review the consequences of elliptical-galaxy formation on a downsizing timescale, which implies a significant extragalactic foreground released at $15< z < 20$. Under conservative assumptions, this foreground contributes at least 1.4 percent of the observed CMB energy density and has not been included in CMB analyses. {\it It follows that the reported $10^{-5}$-level fluctuations is introduced by overly aggressive foreground cleaning.}}

\FullConference{Corfu Summer Institute, Tensions in Cosmology, Sept. 2-8, 2025
}

\begin{document}
\maketitle

\section{Introduction}
\label{sec:introd}

The observed cosmic microwave background (CMB) constitutes the central pillar of modern cosmology and counts as the primary evidence for a Hot Big Bang 
\citep[e.g.,][]{Staggs+2018, PlanckRev2020}. It is understood 
that the CMB stems from recombination 
when the hot plasma had cooled sufficiently at a redshift of $z\approx 1100$ making the Universe transparent to radiation.  The observational establishment of this relic radiation came with the space missions COBE (1989--1993, WMAP (2001--2010) and Planck (2009--2013). The fluctuations in temperature and gas density of $\delta T/T \approx \delta \rho/\rho \approx 10^{-5}$
installed at recombination from the prior evolution out of the inflationary epoch
evolved over cosmic history to the present-day cosmological structures. It is generally well accepted that the current standard model of cosmology (SMoC) based on Einstein's theory of general relativity (GR) augmented by a  brief very early inflationary epoch, a major cold or warm dark matter (C/WDM) content and an increasing dark energy (in terms of the cosmological constant $\Lambda$) contribution is needed to account for the CMB properties. The temperature and polarization angular power spectra constructed by precise measurements, data reduction and analysis  allow an accurate extraction of the six basic parameters defining this model (table~6 in \cite{PlanckRev2020}). A vast literature accounts for the very major research effort that has gone into establishing the theoretical basis for understanding the CMB within this standard model. While the results achieved are impressive, the SMoC has been showing very major problems if not catastrophic failures (e.g. \cite{PeriSkara2022, DiValentino2022, DiValentino2022b, Kroupa+2023, Kumar+2023}). The Hubble Tension is currently seen to be perhaps the major such failure (e.g. \cite{Haslbauer+2020,DiValentino+2021d,Abdalla+2022, BanikSamaras2024, Mazurenko+2025}), but a significantly greater failure comes in the Chandrasekhar dynamical friction test falsifying the existence of dark matter particles with more than five sigma confidence \citep{OehmKroupa2024, HernandezKroupa2025}. 

If the SMoC is indeed the wrong theory of cosmology, then what role does the CMB play in this problem? Stress between the observationally derived CMB and its power spectra and the expectations of the best-fit SMoC in terms of the $\Lambda$CDM model have been noted \cite{Schwarz+2016, DiValentino2022, DiValentino2022b} and also discussed in view of the overall performance of the SMoC \cite{Kroupa+2023}. In view of further recent progress on this matter, the following questions are raised in the respective sections:\\[-5mm]
\begin{enumerate}
\item 
Sec.~\ref{sec:uniqueness}: Model uniqueness: Is the SMoC the only model able to reproduce the currently published properties of the CMB; that is, do the currently published observationally derived constraints on the CMB and it's power spectra rule out any non-SMoC?\\[-8mm]

\item 
Sec.~\ref{sec:signature}: The handwriting of God? (George Smoot) [OR] The Holy Grail of cosmology? (Michael Turner): Does every cosmological model need to reproduce the currently published CMB and its power spectra; that is, are these empirical constraints truly universally valid for all models of cosmology?\\[-8mm]

\item 
Sec.~\ref{sec:foreground}: Significant foreground contamination: How much does galaxy formation at $z>10$ affect the CMB? \\[-5mm]

\end{enumerate}

This contribution concisely summarises recently uncovered scientific problems plaguing CMB science.\footnote{A detailed presentation of this material is available on YouTube at 
\url{https://darkmattercrisis.wordpress.com/2025/12/26/112-the-meaning-of-the-cosmic-microwave-background-cmb-for-cosmology-and-the-role-early-galaxies-have-in-this-matter/}.}
The reader is referred for technical details to the cited literature.

\section{Model uniqueness}
\label{sec:uniqueness}

In this section we address the question {\it Is the SMoC the only model able to reproduce the currently published properties of the CMB?}

\subsection{Different models}

Along with the Hubble-Lemaitre expansion as evident in galaxy redshifts\footnote{Cosmic expansion surely counts as one of the very greatest scientific discoveries ever made. Since Carl Wilhelm Wirtz identified this phenomenon earlier than the others usually credited in this context
\cite{Wirtz1922, Wirtz1924, Richtler2024},
the Hubble-Lemaitre law ought to be named the Wirtz-Hubble-Lemaitre law.},
the observed CMB is taught to be the irrefutable evidence, if not proof, that the Universe was born as a Hot Big Bang which created all matter at once (our 'basic assumption'). The parameters underlying the cold-dark-matter based SMoC (such as the $\Lambda$CDM model) have been adjusted in intimate conjunction with the reduction of the raw CMB data (see Sec.~\ref{sec:foreground}) to produce the observationally constrained CMB and its power spectra as published in the final data release analysis by the Planck team in 2018, and the fit is excellent \cite{PlanckRev2020}. Warm-dark matter models pose a slight variation attempting to address the small-scale problems dark-matter-based structure formation has been facing, but are essentially indistinguishable from the cold-dark-matter based SMoC (\cite{Kroupa+2023} and references therein).  But with the falsification of dark matter particles using the Chandrasekhar dynamical friction test \cite{OehmKroupa2024,HernandezKroupa2025}, all such dark-matter-based models have become irrelevant\footnote{Even though still being argued to be a model of success, Michael Turner does point out the need for the SMoC's replacement \cite{Turner2026}).}. 

It is therefore important to study if models that do not assume cold or warm dark matter particles can nevertheless be consistent with the currently published CMB and its power spectra. Dark matter was originally introduced with the aim of providing additional gravitational potential to account for the dynamical and lensing masses of galaxy clusters and for the rotation curves of galaxies, but it is also needed to fit GR to the observationally-deduced peaks of the CMB temperature power spectrum and to allow structures to form in the expanding Universe as it's observed baryonic matter content alone does not suffice (under the 'basic assumption' above). Without dark matter, this extra potential can only be accounted for by an effectively stronger gravitational force law. The most successful and tested such approach is that following the discovery of a ubiquitous acceleration scale, $a_0 \approx3.8\,$pc/Myr$^2$, in~1983 and~1984 by Milgrom and Bekenstein \cite{Milgrom1983a, Milgrom1983b, Milgrom1983c, BekensteinMilgrom1984}; we refer here to the implied gravitational dynamics as "Milgromian dynamics" (MOND) or Milgromian gravitation which supersedes Newtonian gravitation.  While originally being a formulation based on a non-relativistic Lagrangian \cite{BekensteinMilgrom1984}\footnote{In this regime Newtonian gravitational dynamics is described by the Poisson equation which corresponds to the $p=2$ Laplacian. In the AQUAL formulation of MOND the field equation changes in the very-weak field limit to the $p=3$ Laplacian which is synonymous with isolated point masses generating logarithmic potentials around them (see also \cite{Scherer+2025}).}, relativistic formulations of MOND have been pioneered by Bekenstein in 2004 \cite{Bekenstein2004} and further developed (e.g. Skordis \& Sloznik 2021 \cite{SkordisZlosnik2021}). The reader is referred to the available reviews of this non-linear theory of gravitation in the non-relativistic limit \cite{FamaeyMcGaugh2012, Milgrom2014, Merritt2020, BanikZhao2022}. Recent research has verified the MOND-predicted asymmetry of the tidal tails of open star clusters on a scale of a few~pc \cite{Kroupa+2022, Kroupa+2024},\footnote{Detecting the members of tidal tails is subject to bias such that more work is needed and underway to improve the astrometric constraints.} that the gravitational potential around isolated galaxies is logarithmic out to a~Mpc \cite{Mistele+2024}, and that the purely baryonic masses of nearby galaxy clusters are consistent with MOND on a scale of a few Mpc \cite{Zhang+2026}. MOND accounts for the dynamical behavior observed in systems that can be approximated with continuum dynamics, that is, in collisionless systems that have a median Newtonian two-body relaxation time comparable to a Hubble time \cite{EappenKroupa2026}. For systems in which few-body dynamics becomes important the mean-field formulation of MOND breaks down but the discrete-few-body dynamics is mathematically not well understood due to the non-linearity of Milgromian dynamics -- it is unclear how the gravitational forces are to be added \cite{Pflamm-Altenburg2025a, Pflamm-Altenburg2025b}. In this regime astrophysical degeneracies (e.g. metallicity/age/luminosity in binary stellar systems) as well as mass-values (e.g. the masses of the outer planets have been obtained assuming Newtonian dynamics to be valid, Pflamm-Altenburg, private communication) lead to systematic biases that affect our current assessment of Milgromian or Newtonian dynamics being valid in very wide binaries and the outer Solar system. It may thus not be surprising that the conclusions whether Milgromian or Newtonian dynamics is valid in this regime remain unclear.\footnote{Concerning very wide binaries 
Banik et al. (2024 \cite{Banik+2024}
report the astrometric data to imply Newtonian dynamics while Hernandez et al. (2022 \cite{Hernandez+2022}, 2024 \cite{Hernandez+2024}) and Chae (2022 \cite{Chae2023}, 2026 \cite{Chae2026}) find a distinct Milgromian signature. Concerning the outer Solar system, Pauco \& Klavicka (2016 \cite{PaucoKlavicka2016}), Pauco (2017 \cite{Pauco2017}) and Brown \& Marthur (2023 \cite{BrownMathur2023}) report possible Milgromian effects while Vokrouhlicky et al. (2024 \cite{Vokrouhlicky+2024}) report Newtonian behaviour.}

Given the success of MOND to account for collisionless systems, MOND-based cosmological models have been studied in view of the currently published CMB and its power spectra, and McGaugh (1999 \cite{McGaugh1999a}) predicted for the first time, not only that the CMB temperature power spectrum must have two major peaks, but also used these as a test how MOND-based (i.e. a purely baryonic-matter) cosmology compares to the cold/warm-dark-matter  based models in terms of the ratio of the amplitudes of these first two peaks.  
Under our 'basic assumption' and that Milgrom's acceleration scale $a_0$ remains constant with time, the early cosmological epoch is in the strong-gravitational-field regime such that the effective enhancement to gravitation through MOND is negligible. The later measurements of these two peaks with the WMAP mission falsified this purely baryonic model Universe. 

\subsection{The $\nu$HDM models}

A first simple version of MOND-based cosmology was developed by adopting the near-to-exact expansion history as in the SMoC keeping the Friedman-Lemaitre-Robertson-Walker (FLRW) metric untouched and replacing the cold dark matter component with an 11~eV sterile neutrino component which effectively constitutes hot dark matter (HDM). This $\nu$HDM model 
is consistent with the WMAP measurements of the CMB (and notably the relative heights of the peaks, Angus 2009 \cite{Angus2009}) and accounts for flat rotation curves because the sterile neutrinos cannot condense in galaxy-scale potentials. The mass-discrepancy which arises in MOND with lensing masses requiring about twice as much matter than is observed in galaxy clusters is also solved in this $\nu$HDM framework because the sterile neutrinos condense as halos of hot dark matter around sufficiently massive baryonic components \cite{Angus+2010, Wittenburg+2023}.\footnote{\label{ftn:nuHDMfailure} This has meanwhile become a failure of the $\nu$HDM model because it has only recently been realised that galaxy clusters have, due to their high metallicities, a significant population of stellar-mass black holes and neutron stars. These stem from the massive stars needed to provide the metal enrichment. Galaxy clusters therefore do not have a missing-mass problem in MOND and the $\nu$HDM (and the opt-$\nu$HDM -- see below) model would therefore lead to galaxy clusters of a given baryonic mass having about two times too much dark mass (stellar remnants plus the sterile neutrinos) than allowed by weak lensing constraints \cite{Zhang+2026}.} The $\nu$HDM model has therefore been very attractive such that hydrodynamical \cite{Wittenburg+2023} and pure $N$-body structure formation simulations have been performed \cite{Katz+2013, Russell+2026}. 
All of these simulations indicate that this model fails to reproduce observed structures in that the most massive clusters formed are significantly too massive ($\approx 10^{17}\,M_\odot$) compared to observed galaxy clusters. This is emphasised by Russel et al. (2026 \cite{Russell+2026}) whose collisionless $N-$body simulations of large-scale structure formation appear to rule out the $\nu$HDM model because it creates significantly too fast bulk flows and significantly too massive galaxy clusters \cite{Russell+2026}, a problem already noted by \cite{Wittenburg+2023}. Also, observable structure formation appears to begin too late (similarly to the SMoC as shown by Samaras \cite{SamarasPhD2025}.). 

\begin{figure*}
    \centering
    \vspace{-15 pt}
    \includegraphics[width=0.9\columnwidth]{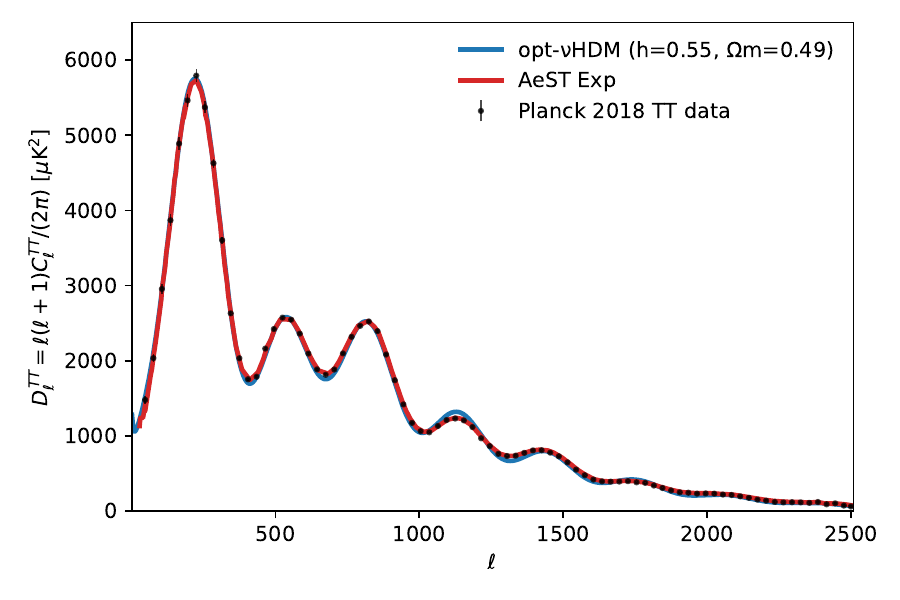}
    \vspace{-13 pt}
    \caption{The CMP temperature fluctuation power spectrum. The black data points are the final release Planck 2018 data (the "Planck-CMB data", \cite{Planck2018}). The blue curve shows the opt-$\nu$HDM model (Samaras et al. 2025 \cite{Samaras+2025}, note that full optimisation of this fit has not been sought because the baryon fraction was kept constant at the SMoC value) and the red curve is the new relativistic MOND AeST model (Skordis \& Zlosnik 2021 \cite{SkordisZlosnik2021}). The $\Lambda$CDM model (the SMoC) is not explicitly shown here but is indistinguishable from the Planck data.  This figure demonstrates that the final Planck-CMB data can be reproduced by at least three very different cosmological models.  
    }
    \label{fig:CMBpower}
\end{figure*}

One possible way out is to fit the model parameters (the same as in the SMoC with the C/WDM being replaced by HDM) to the Planck-CMB data to obtain an optimised version (the opt-$\nu$HDM model, Samaras et al. 2025 \cite{Samaras+2025}). This opt-$\nu$HDM model with fitted cosmological parameters serves as an example of a formulation very different to the SMoC but nevertheless fitting the Planck-CMB data exceedingly well. Fig.~\ref{fig:CMBpower} demonstrates this. The technical details can be found in \cite{Samaras+2025}. It suffices to point out here that this model has, compared to the SMoC's matter content, more mass in the sterile neutrinos but the same mass in baryons (in total $\Omega_{\rm m} \approx 0.5$) and a smaller Hubble-Lemaitre constant ($H_0/({\rm km/s/Mpc}) = h\,100 = 55.6$) and consequently is less dominated by dark energy than the SMoC. Not only does it fit the Planck-CMB data very well, but also automatically solves the small-scale problems the community has been addressing since about 30\,yr (e.g. \cite{Kroupa+2010, FamaeyMcGaugh2012, BanikZhao2022, Kroupa+2023}). This opt-$\nu$HDM model thus constitutes an improvement over the SMoC, may be pointing into the right theoretical direction (no cold or warm dark matter but MOND) but may not yet be realistic (see footnote~\ref{ftn:nuHDMfailure}). 

As a caveat concerning discarding the opt-$\nu$HDM model, it needs to be emphasized that the observed Universe appears to be highly inhomogeneous up to a scale of few~Gpc \cite{Haslbauer+2023} and possibly over its entire expanse (Sec.~\ref{sec:asym}) and that the most-massive galaxy clusters may be as massive as predicted by this model. As these would be located at a co-moving distance of about 5~Gpc away from us we would not be able to observe their present-day masses, because they would be caught at a time (corresponding to $z\approx2$) when their masses are assembling. Likewise, the bulk-flow velocities observationally deduced from the local Universe may not suffice to discard the model since major bulk flows need time to develop whereas the distant Universe appears time-retarded to the terrestrial observer. The masses of galaxy clusters (footnote~\ref{ftn:nuHDMfailure}) and the very early formation of galaxies (Sec.~\ref{sec:foreground}) may however be failing points. 

\subsection{The AeST model}

A different approach to the above more pragmatical one is to develop a cosmological model based on a new general relativistic theory that caters for the strong-GR, the weak-GR (Newtonian) and the ultra-weak-field (Milgromian) regimes. The work of Bekenstein (2004 \cite{Bekenstein2004}) constitutes a breakthrough as it for the first time demonstrated that such theories can be mathematically formulated. Instead of adding an exotic (dark matter) component which has a largely undefined number of degrees of freedom (particle masses, possibly different dark matter species, interaction cross sections, also in the dark sector), this formulation introduces a scalar, a vector and a tensor field. The detectability of these fields and their role in terms of structure formation remain unclear. The scalar field is relevant for the gravitational potential energy, the vector field for the large-scale motion of matter, and the tensor field is relevant for the curvature of spacetime. These fields work together to account for the baryonic acoustic oscillations as well as the dynamics in galaxies and galaxy clusters.

The original formulation by Bekenstein was falsified by the empirical result that electromagnetic and gravitational waves propagate with the same speed $c$ \cite{Abelson2022}, but Skordis \& Zlosnik (2021 \cite{SkordisZlosnik2021}) remedied this problem with a new relativistic theory of MOND, referred to as the Aether Scalar Tensor (AeST) theory. The AeST theory is explicitly formulated to reproduce the same expansion history as the SMoC and to also reproduce the currently published Planck-CMB constraints. The agreement with the Planck-CMB data is excellent, as is evident in Fig.~\ref{fig:CMBpower}. Weak lensing data around isolated galaxies appear to be compliant with this theory \cite{Mistele+2023}.
It can be argued that not much is gained since the introduced fields
can be interpreted as exotic dark matter particle fields. However, their equations of motion differ from those of dark matter particles that are ballistic, but the effect on structure formation remains to be studied (e.g. Thomas et al. 2023 \cite{Thomas+2023}).

\subsection{Conclusion to Sec.~\ref{sec:uniqueness}: Lesson~I}

The Planck-CMB data can be fitted equally well by very different cosmological theories ($\Lambda$CDM and by implication $\Lambda$WDM, opt-$\nu$HDM, AeST) such that it is not correct to argue that the SMoC is unique in accounting for the high-precision measurements of the surface of last scattering. This is in fact a demonstration of the fundamental scientific method: data never prove a theory since multiple theories can account for the same data. Empirical data can only falsify a theory.

\section{The handwriting of God? (George Smoot) The Holy Grail of cosmology? (Michael Turner)}
\label{sec:signature}

This section's title draws on the rhetoric that accompanied the early CMB anisotropy results, in which the CMB was presented as one of the highest observational achievements in cosmology \citep{Smoot2006Biographical}. In this section we address the question {\it Does every cosmological model need to reproduce the currently published CMB power spectra? That is, are these Planck-CMB data invariantly valid for every cosmological model?}

As an explicit example, studies of structure formation in the opt-$\nu$HDM and the AeST models (Sec.~\ref{sec:uniqueness}) have been adopting the currently published Planck CMB constraints as initial conditions at $z\approx 1100$ with the AeST model being explicitly formulated to precisely reproduce these. Researchers have thus indeed assumed that the currently published Planck-CMB data need to be fitted by every conceived cosmological theory. 

\subsection{The CMB is not observable, it must be calculated from the raw data assuming a cosmological model}
\label{sec:unobservable}

The cosmologically relevant CMB constraints, in the sense that the CMB is the relic radiation field from the surface of last scattering that occurred at $z\approx 1100$ as the photosphere of the Hot Big Bang, are not observable. They need to be calculated from the radiation field which the Solar system is immersed in. To extract the relic radiation field from this photon field it is necessary to apply four corrections that lead to fluctuations in the photon field even if there were no fluctuations at $z\approx1100$ (such that the power spectra would be flat):\footnote{An award-winning exposition of this problem with a list of relevant research papers can be found on \url{https://cosmologyscience.com/cosblog/cosmic-microwave-angular-resolution-surprise/}.}\\[-8mm]
\begin{enumerate}
\item Identify and remove all foreground contributions (Solar system, the Milky Way, nearby and distant galaxies).  \\[-8mm]

\item Correct for the changing propagation directions of the photons through weak lensing on cosmological structures that evolve as time proceeds and the Universe expands.\\[-8mm]

\item Correct for the changing energies of the photons through scattering on electrons which accumulate within the evolving potential wells of the evolving cosmological structures (the Sunyaev-Zeldovich effect).\\[-8mm].

\item Correct for the changing energies of the photons as these traverse potential wells that evolve as the photons traverse them (integrated and non-integrated Sachs-Wolfe effect).\\[-8mm]

\end{enumerate}

\noindent The first correction can be accounted for using known sources and statistically assuming galaxies begin to emit photons at a redshift near $z\approx5$ which is the expectation in the SMoC \cite{Haslbauer+2022b, McGaugh+2024}. The other three corrections can only be applied statistically because structure growth is not observable given that the majority of matter is dark in the SMoC. The raw Planck data were processed and reduced assuming the SMoC to be the valid model of structure formation. But other models lead to different rates of structure formation and also to significantly larger voids and under- and over-densities. For example, the KBC void, which today has a radius near 300~Mpc cannot form in the SMoC but it is clearly evident in tracers across the entire electromagnetc spectrum  \cite{Haslbauer+2020,Mazurenko+2024,Mazurenko+2025,BanikKalaitzidis2025}. The Universe appears to be significantly inhomogeneous on a~Gpc (Lopez \& Clowes 2025 \cite{LopezClowes2025}) and a few~Gpc scales (Haslbauer et al. 2023 \cite{Haslbauer+2023}), and observations with the James Webb Space Telescope (JWST) have shown massive ($\approx10^9\,M_\odot$ in baryons) galaxies to have formed already at $z\approx15$. The significantly more enhanced structure growth at all scales evident in the observational data than allowed by the SMoC (which underlies the above four corrections in the currently published Planck-CMB data) means that the corrections will be different. The true CMB power spectra thus remain unknown.

Indeed, the SMoC faces problems with fitting the CMB on large angular scales \cite{Schwarz+2016, Ivanov+2020, ODwyer+2020} which might be related to enhanced growth of structure
on scales laerger than $100\,$Mpc. McGaugh (2024, \cite{McGaugh2024}) points out that the not-expected (based on  the SMoC) but observed early appearance of massive galaxies at $z>7$ affects the CMB through weak lensing of CMB photons, there being more power at small angular scales than expected, possibly pushing the cosmological parameter fits away from those obtained previously with the WMAP mission. The effects of large-scale inhomogeneities on the interpretation of the CMB properties has been discussed by  C{\'e}l{\'e}rier (2024 \cite{Marie-Noelle2024}).
An observer in an underdense region of the Universe will see a redshifted CMB such that the true CMB is hotter. For the KBC void, which solves the Hubble tension, this problem is however negligible   \cite{Haslbauer+2019b}.  The effect may be non-negligible if the Universe has inhomogeneities on the scale of a few~Gpc and density amplitudes of a factor of a hundred as is indicated to be the case by observations \cite{Haslbauer+2023}.

The problem that the currently published Planck-CMB data cannot be "The handwriting of God", i.e. a necessary extremely tight constraint to be fulfilled by every conceived cosmological model, was for the first time raised by Samaras et al. (2025  \cite{Samaras+2025}) and Gjergo \& Kroupa (2025 \cite{GjergoKroupa2025}). It follows that, in order to test any particular model of cosmology, the CMB and its power spectra need to be recalculated from the raw microwave flux data for each model separately.

\subsection{Conclusion to Sec.~\ref{sec:signature}: Lesson~II}

The currently published Planck-CMB data are strictly only valid as precise cosmological constraints for the SMoC which however is an incorrect model of the real Universe. Force-fitting any other Hot-Big-Bang cosmological model to comply with these data likely leads to an unphysical model of cosmology. At the moment we do not know how large the error is when, for example, forcing the AeST model into the current Planck-CMB data. The calculations have not yet been done how the four corrections, when applied for example using the opt-$\nu$HDM model to reduce the raw Planck data to obtain the physically correct self-consistent initial conditions for the opt-$\nu$HDM model, affect the power spectra.  

The correct procedure for this must be to (a)~start with a first-guess for the values of the parameters of a Hot Big Bang cosmological model and perform simulations of structure-formation and evolution (Samaras \& Kroupa, in prep.). (b)~Apply these to a first estimate of the four corrections (1-4) above to obtain a first estimate of the CMB power spectra from the raw Planck data. (c) With these newly calculated properties of the CMB, refit the cosmological model to obtain improved new values of the parameters that define the cosmological model.  Then (d) perform a new set of simulations of structure formation and evolution based on the new parameters. (e) This allows an improved estimate of the four corrections to extract a new estimate of the CMB properties from the raw data to (f) allow a new fit of the cosmological parameters with which new cosmological structure formation simulations are performed. This procedure needs to be iterated until the fitted parameters and the computed CMB properties converge to stable values. If no convergence is possible then the cosmological model might have to be discarded for internal inconsistencies.


\section{Significant foreground contamination}
\label{sec:foreground}

In this section we address the question {\it Is the observed CMB affected by galaxy formation at $z > 10$?}

\subsection{The four corrections}

As already stated in Sec.~\ref{sec:signature}, McGaugh (2024 \cite{McGaugh2024}) pointed out that the observed but not expected (based on the SMoC) appearance of massive galaxies at $z>7$ is likely to affect weak lensing corrections of the Planck-CMB power spectra at small angular scales that may be responsible for driving the Planck-CMB data parameter fit away from the previously obtained WMAP fit. Also, evidence for large-angular-scale problems between the Planck-CMB data and the SMoC have been noted in Sec.~\ref{sec:signature}. But in addition to distortions of the CMB power spectra due to incorrect corrections for weak lensing, the Sunyaev-Zeldovich and Sachs-Wolfe effects, Gjergo \& Kroupa (2025 \cite{GjergoKroupa2025}) have pointed out that the CMB monopole, i.e. the energy density of photons, may contain a significant contamination by forming massive galaxies at $z>15$. 


\subsection{Massive elliptical galaxies formed early and rapidly}
\label{sec:massiveEs}

It had been known since many decades that elliptical galaxies with stellar masses $M_*>10^{10}\,M_\odot$ mostly consist of old stellar populations. Notably, Matteucci (1994 \cite{Matteucci1994}) had already deduced that elliptical galaxies needed to form within timescales of a few hundred~Myr with top-heavy galaxy-wide stellar initial mass functions (gwIMFs). This seminal deduction was based on their super-Solar metal abundances and super-Solar $\alpha$-element over iron ([$\alpha$/Fe]) ratios (Weiss, Peletier \& Matteucci 1995 \cite{Weiss+1995}; Gibson \& Matteucci 1997 \cite{GibsonMatteucci1997}; see also Vazdekis et al. (1997 \cite{Vazdekis+1997} and the discussion and references within Kroupa et al. (2020 \cite{Kroupa+2020}).

The work by Thomas et al. (1999 \cite{Thomas+1999}, 2005 \cite{Thomas+2005}) based on spectroscopic $\alpha$-elemental abundances, by McDermid et al. (2015 \cite{McDermid+2015}) based on stellar population synthesis and by Yan et al. (2021 \cite{Yan+2021}) based on chemical evolution calculations has demonstrated robustly that the formation time is earlier and quicker the larger $M_*$ is. This is well known as the downsizing problem in the SMoC which predicted that more massive galaxies take longer to assemble due to more time being needed for sufficient mergers to occur. A detailed comparison between the spread of stellar ages and their formation times falsifies a merger-origin of the stellar populations in elliptical galaxies with much more than 5$\sigma$ confidence (Eappen et al. 2022 \cite{Eappen+2022}), a failure of the SMoC already emphasised by Thomas et al. (1999 \cite{Thomas+1999}) and Pipino \& Matteucci (2008 \cite{PipinoMatteucci2008}). 

\begin{figure*}
    \centering
    \vspace{0 pt}
    \includegraphics[width=0.65\columnwidth, angle=90]{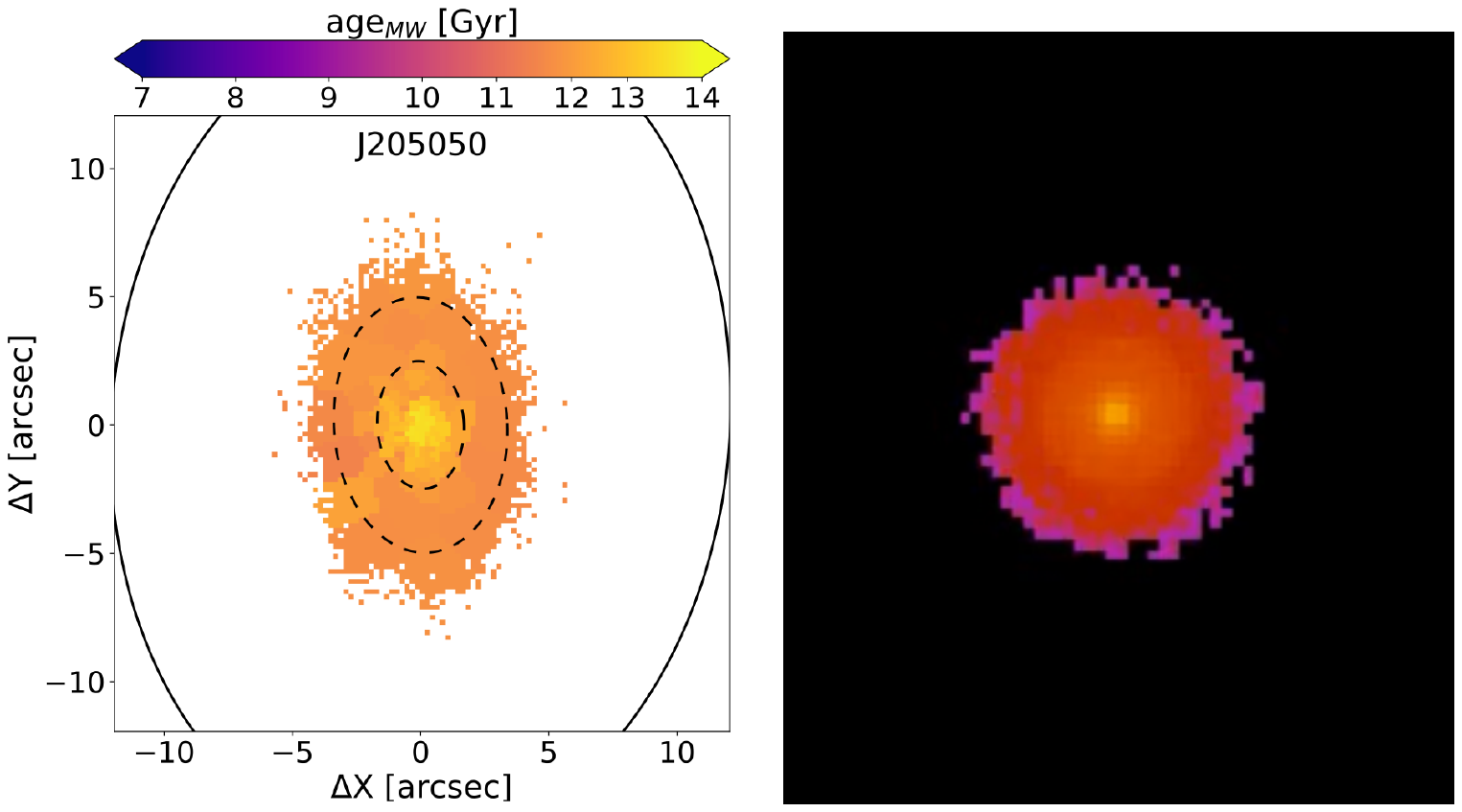}
      \vspace{-40 pt}
    \caption{Stellar ages in a nearby (left panel) and simulated (right panel) elliptical galaxy.  The {\it left panel} is for galaxy~J205050 from Jegathesaan et al. (2025 \cite[their fig.~6]{Jegatheesan+2025}). The color coding shows the mass-weighted ages of the stars; note the ages $>14\,$Gyr -- see text for further discussion. 
    The {\it right panel} shows the stellar particles formed in the simulation of a collapsing $10^{11}\,M_\odot$ pre-galactic gas cloud by Wittenburg et al. (2020 \cite{Wittenburg+2020}) and Eappen et al. (2022 \cite{Eappen+2022}). This model galaxy quenches after about 500~Myr and remains at the shown morphology for 10\,Gyr. The star-burst begins at the center and the new stellar particles form from the collapsing gas at increasing radii such that the oldest stellar particles remain near the center-most region, as is the case in the left panel. The physical scale is similar in both cases with the inner and outer dashed ellipse in the left panel being the effective radius (2.49\,kpc) and twice its value, respectively.}
    \label{fig:jegatheesan}
\end{figure*}

That massive elliptical galaxies formed extremely early (14--16~Gyr ago) and very rapidly (within a Gyr) is meanwhile an established observational fact confirmed by a number of independently working groups. Based on their independent survey Thomas et al. (2010 \cite{Thomas+2010}, their fig.~2) find that the most massive elliptical galaxies with line-of-sight velocity dispersions larger than 120\,km/s reach stellar-population-synthesis ages of about 16~Gyr with super-Solar metal abundances. The same result is evident also in the independent work of Li et al. 2018 \cite{Li+2018}, their fig.~5). The most recent observational work on this problem by Jegatheesan et al. (2025 \cite{Jegatheesan+2025}), who document with high-resolution observations the spatial distribution of stellar ages and metallicities in a number of nearby elliptical galaxies, shows their centers to consist of stars that, according to the stellar population synthesis results, are older than the nominal age of the Universe (13.8~Gyr). The populations are extremely old throughout the galaxies and reach significantly-super-Solar metal bundances. Fig.~\ref{fig:jegatheesan} demonstrates an example. 

These and many other observational results compellingly show that elliptical galaxies formed their bulk stellar populations through the free fall monolithic collapses of pre-galactic gas clouds, since only the free-fall time scale is compatible with the extremely rapid (Gyr or shorter) formation times deduced from the stellar population properties. The different groups consistently find the elliptical galaxies to have formed 14--16~Gyr ago, suggesting that the age of the Universe is larger than 14--16~Gyr, unless our understanding of long-lived stars is wrong. The monolithic collapses of primordial pre-galactic gas clouds is entirely impossible in the SMoC because the model requires a highly homogeneous dark-matter dominated matter distribution and cannot accommodate star formation 14--16~Gyr ago. The research on elliptical galaxies thus very clearly shows that only models of cosmology are viable that have sufficient density contrasts 14--16~Gyr ago with enhanced gravitation (since dark matter does not exist by the Chandrasekhar dynamical friction test  \cite{HernandezKroupa2025}) to allow such monolithic collapses to have occurred throughout the Universe. 

The result of a pre-galactic gas cloud collapse simulation in MOND by Wittenburg et al. (2020 \cite{Wittenburg+2020}) and Eappen et al. (2022 \cite{Eappen+2022}) is shown in Fig.~\ref{fig:jegatheesan} for a comparison to an observed elliptical galaxy by Jegatheesan et al. (2025 \cite{Jegatheesan+2025}). The chemical evolution modelling of such galaxies by Yan et al. (2021 \cite{Yan+2021}) finds that in order to simultaneously reproduce $M_*$, the metallicity $Z$ and [$\alpha$/Fe], the galaxies need to form before supernova type 1a significantly contribute iron and with top-heavy gwIMFs. The implied star-formation rates ($SFR$) these galaxies had while forming were immense ($10^3<SFR/\left(M_\odot\,{\rm yr}^{-1}\right)<10^4$), with such levels of $SFR$ being observed for quasar host galaxies with ALMA at $z\approx 4.8$ \cite[their fig.9]{Nguyen+2020}.\footnote{Such high values of $SFR$ are also implied in the formation of super-massive black holes together with their host galaxies \cite{Kroupa+2020}.} 

This suggests that the real Universe went through a very early epoch of hyper-star bursts occurring throughout it leaving the presently observed elliptical galaxies as remnants of this brief radiant epoch.

\subsection{The radiation field emitted by the early hyper-star bursts}

The observed properties of elliptical galaxies in conjunction with our knowledge of stellar populations thus compellingly imply that the real Universe went through an early epoch of hyper-star bursts at the co-moving locations where the present-day elliptical galaxies are. The calculation by Gjergo \& Kroupa  (2025 \cite{GjergoKroupa2025}) of the luminosity evolution of massive elliptical galaxies with present-day bolometric luminosities near $10^{12}\,L_\odot$ shows these to have reached peak bolometric luminosities of $10^{15}\,L_\odot$  during their formation as a result of the nucleosynthesis to the significantly-super-Solar metallicities they are observed to have (e.g. fig.~4 in  Thomas et al. 2010 \cite{Thomas+2010}; fig.~5 in Li et al. \cite{Li+2018}). It is also well known that star-forming, and in particular star-bursting galaxies thermalise their stellar emission through dust processing and thus emit most of their radiation in the infrared near a wavelength of about $170\,\mu$m (Milky-Way type galaxies) or $60\,\mu$m (ultra-luminous infrared galaxies, ULIRGs) as black-body-like emitters (e.g. fig.~2 in Iyer et al. 2025 \cite{Iyer+2025}). The peak in the black-body-like emission spectrum depends on the dust temperature and thus also on the $SFR$. If we take $50\,$K to be a typical dust temperature and the hyper-star bursts to be occurring at $z=17$, then they would, in our $z=0$ rest frame, appear as 2.77~K emitters. Thus, the rest-frame dust peak emission near $100\,\mu$m $=0.1\,$mm of a hyper-star burst shifts when observed at $z\approx17$ to a maximum near $1.9\,$mm where the observed CMB spectral energy distribution peaks. 

Given that we know the present-day locations of the massive elliptical galaxies (they have separations of about 15~Mpc), their co-moving locations and thus number per co-moving Gpc$^3$ are known as well. This allowed Gjergo \& Kroupa (2025 \cite{GjergoKroupa2025}) to calculate the photon energy density released from the forming elliptical galaxies and its evolution as the Universe expanded. The time-reversed shrinking Universe implies the physical distances to have reached about 800~kpc at $z\approx 17$. The pre-galactic-gas-cloud collapse simulations by Eappen et al. (2022 \cite{Eappen+2022}) inform us that in order to reproduce the downsizing times, namely the dependency of the formation time-scale of an elliptical galaxy on $M_*$, the clouds had to have had initial radii near 400~kpc independently of their mass.

Three interesting ``coincidences'' emanating from this work are (i)~that the earliest the gas clouds could have collapsed is when their radii touched, i.e. at $z\approx17$ taking the present-day observed separations of such galaxies (about 15~Mpc). Also, (ii)~the black-body-like 50~K-warm dust emission emitted at this redshift peaks today at 2.77~K, and (iii)~the photon energy density emitted by the forming elliptical galaxies corresponds to the energy density of the observed CMB field at the present time (Gjergo \& Kroupa 2025 \cite{GjergoKroupa2025}). 

These three ``coincidences'' appear too natural to ignore. If the elliptical galaxies had formed at, say, $z\approx5$, and if their stellar populations formed in-situ, then each galaxy would have radiated the same photon energy with a rest-frame black-body-like dust emission peak near 50~K since the observed metallicities needed to be synthesized by massive stars in the galaxies. But the matter density evolves with cosmic time, $t$, as $\propto a(t)^{-3} = \left(1/(1+z(t))\right)^{-3}$, while the photons lose energy through redshift as $\propto 1/a(t)$. This means that a formation redshift near $z=5$ would imply an extragalactic microwave background that would have a $1/a$ times higher energy density than the currently observed CMB and a black-body-like background with a temperature $T_{\rm em}\approx 8\,$K since the black body temperature an observer at $z=0$ receives is $T_{\rm obs} = T_{\rm em}\,a(t)$. Such a background does not exist.

But even if the high metallicities observed in elliptical galaxies had come from prior star formation from which the present-day ellipticals assembled hierarchically, the same problem with the radiation field would persevere because the same amount of metals had to have been synthesized in massive stars.

\begin{figure*}
    \centering
    \vspace{0 pt}
    \includegraphics[width=0.8\columnwidth]{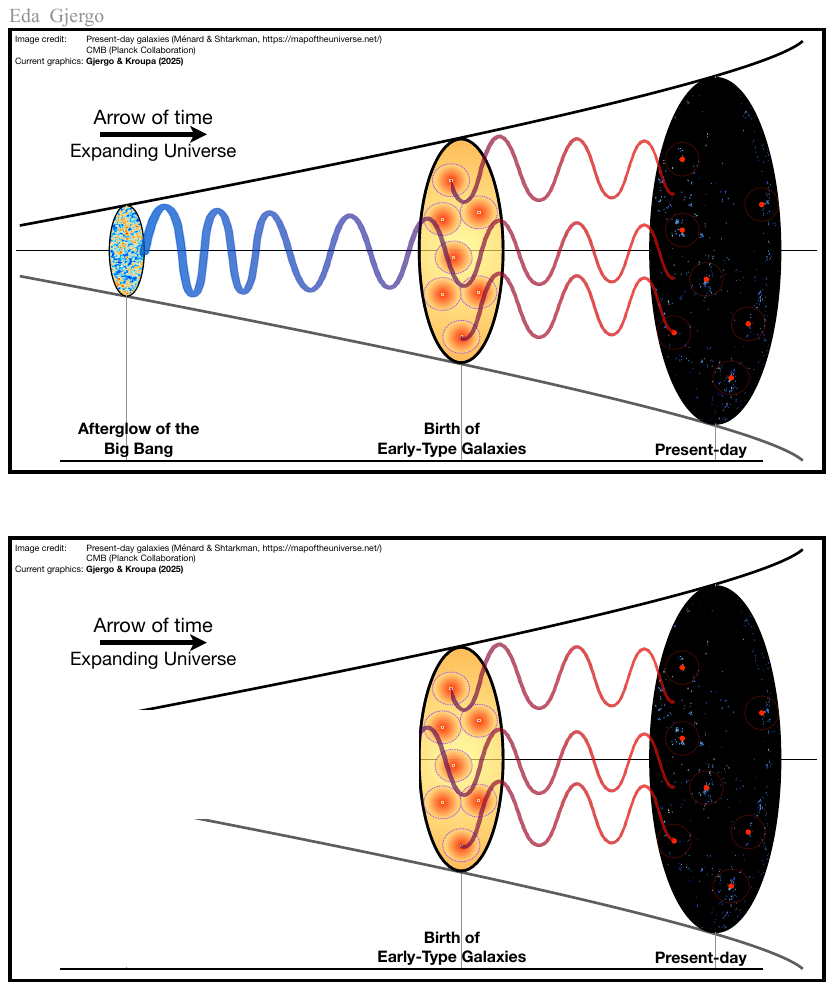}
  \vspace{-115 pt}
    \caption{
      Cartoon based on Gjergo \& Kroupa (2025 \cite{GjergoKroupa2025}) demonstrating in the {\it upper panel} how the emission at $z\approx1100$ from the surface of recombination (the surface of last scattering or the photosphere of the Hot Big Bang or "Afterglow of the Big Bang") is overwritten by the emission from the birth of early-type galaxies at $z\approx 17$. Touching pre-galactic gas clouds collapse monolithically and the formed elliptical galaxies expand from each other as the Universe expands. At the present-day they are but the red embers of once hype-star-bursting objects. The {\it lower panel} demonstrates that at the present-day we would be seeing a CMB even if the afterglow of the Big Bang had not occurred. 
    }
    \label{fig:gjergo}
\end{figure*}

Fig.~\ref{fig:gjergo} demonstrates the situation inferred by Gjergo \& Kroupa (2025 \cite{GjergoKroupa2025}) for the bulk of the massive elliptical galaxies forming at $z\approx 17$. According to the results, if the $z\approx 1100$ relic radiation field were not there, we would nevertheless observe a background radiation field very similar to the detected CMB.  The most conservative assumptions made by Gjergo \& Kroupa \cite{GjergoKroupa2025} show the CMB to be contaminated by at least about 1.4~per cent by forming elliptical galaxies. These are not smoothly distributed\footnote{Gjergo \& Kroupa estimate that on average about 6~forming elliptical galaxies populate one Planck pixel.} such that the $10^{-5}$-level fluctuations in temperature, that constitute the critical initial conditions for cosmological structure formation according to the Planck-CMB data, become entirely unphysical.\footnote{This means that the currently published CMB power spectra contain no physical information.}

The above deductions were made based on the observed physical properties of elliptical galaxies which by themselves question the physical content of the currently published Planck-CMB power spectra.  The JWST observations have indeed uncovered massive galaxies to have formed by $z>10$ that were not expected to exist (\cite{Haslbauer+2022b, McGaugh+2024}). This galactic foreground did not enter into the reduction of the Planck raw data and Li et al. (2026 \cite{Li+2026}) emphasise alone the inadequate modelling of Galactic foreground, let alone of the distant $z>10$ galactic foreground. 

\subsection{Hemispherical asymmetry}
\label{sec:asym}

\begin{figure*}
    \centering
    \vspace{0 pt}
\includegraphics[width=0.85\columnwidth]{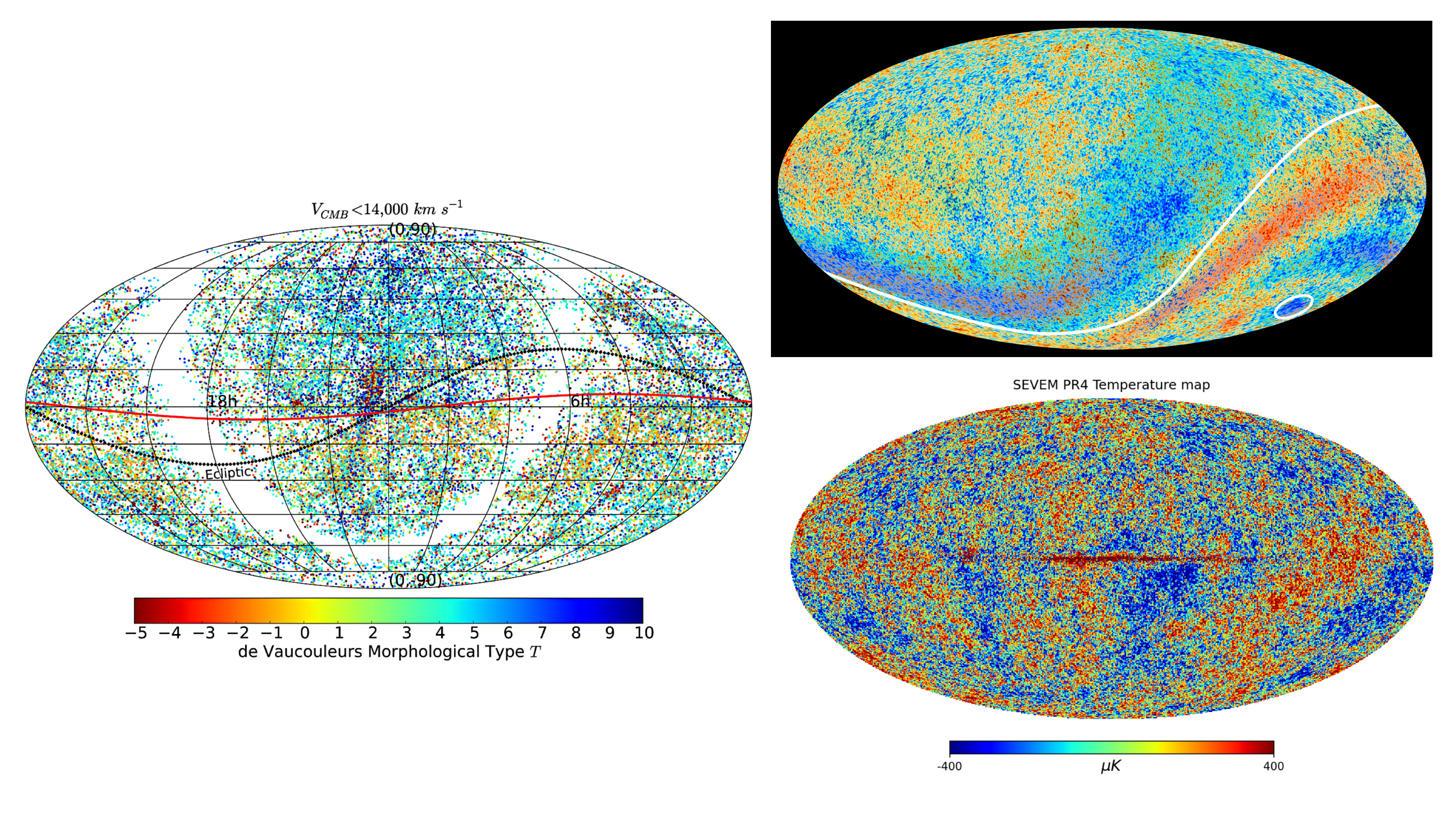}
    \caption{
      The hemispherical asymmetry in the distribution of galaxy morphologies and in the CMB temperature. The blue dots in the {\it left panel} are late-type and the red ones are early-type galaxies in the HyperLeda database with radial velocities relative to the CMB smaller than 14\,000\,km/s (the color bar shows the de Vaucouleurs morphological type; from fig.~6 in Javanmardi \& Kroupa 2017 \cite{JavanmardiKroupa2017}). The black dotted line is the ecliptic and the red line is the plane separating the two hemispheres with the greatest overall difference in the distribution of morphologies.
      The {\it upper right panel} is an enhanced map of the CMB temperature differences with red being warmer and blue colder.  The "cold spot" is highlighted by the white ellipse but not discussed here and the white line delineates the northern and southern ecliptic hemispheres. Credit: ESA and the Planck Collaboration -- see also footnote~\ref{ftn:asymnmetry}. The {\it lower right panel} shows a similar map from Sanyal et al. (2026, \cite[their fig.~1]{Sanyal+2026}). The color bar indicates the CMB temperature difference from the mean in $\mu$K. Note the similarity in sky distribution of the warmer CMB and overabundance of early type galaxies. 
%
%
%
%
    }
    \label{fig:asymmetry}
\end{figure*}

A remarkable additional ``4th coincidence'' exists: The  hemisphere south of the ecliptic contains a much more than 5$\sigma$ overabundance of early-type galaxies noted by Javanmardi \& Kroupa (2017 \cite{JavanmardiKroupa2017}), while the CMB is warmer and has more power in a similar sky region (Eriksen et al. 2004  \cite{Eriksen+2004}; Schwarz et al. 2016 \cite{Schwarz+2016}). The CMB hemispherical power asymmetry is significant but not understood \cite{ODwyer+2020,Sanyal+2026, Carron+2026}. The broad swath on the sky of the significant overdensity of early-type galaxies mapped by \cite[their fig.~6]{JavanmardiKroupa2017} appears to be rather similar to the part of the sky covered by warmer CMB temperatures \cite[their fig.~1]{Sanyal+2026}.\footnote{\label{ftn:asymnmetry}See also 
\url{https://www.esa.int/esearch?q=CMB+hemisphere+asymmetry}.} Fig.~\ref{fig:asymmetry} depicts this situation. 
This coincidence needs to be studied further, but a physical connection between forming early-type galaxies and the CMB would imply a matter inhomogeneity that spans from the redshift of their formation ($z\approx 17$) to the local Universe. 

\subsection{Further constraints on viable cosmological models}

Apart from the contribution to the CMB monopole, the galaxies observed with the JWST at $z>10$ pose severe constraints on any cosmological model. For example, the SMoC is not honestly able to form galaxies as massive (baryonic mass $\approx 10^9\,M_\odot$) as are observed to exist at $z>15$ (\cite{Haslbauer+2022b, McGaugh+2024}). While the SMoC is not at disposition anyway (due to dark matter having been ruled out by the Chandrasekhar dynamical friction test, Sec.~\ref{sec:introd}), the $\nu$HDM and opt-$\nu$HDM models also do not appear to be able to form galaxies at such high redshifts. But this has not yet been conclusively shown because simulations of structure formation based on the correct initial conditions and with sufficiently high resolution are not yet available. The JWST observations require any viable cosmological model to form galaxies by $15<z<z_{\rm u}$, where $z_{\rm u} \approx 20$ if not larger.  The multiple observational findings discussed in Sec.~\ref{sec:massiveEs} according to which the elliptical galaxies formed 14--16~Gyr ago, assuming the current models of low mass stars are correct, pose even stricter conditions on viable cosmological models. 

Any realistic cosmological model must also allow significant matter inhomogeneities to develop by the present, as are observed by the $\approx600\,$Mpc-scale KBC void \cite{Haslbauer+2020}, the $\approx1\,$Gpc-scale "rings" \cite{LopezClowes2025}, and the over-density at co-moving distance of $\approx 5\,$Gpc \cite{Haslbauer+2023}, as well as the CMB hemispherical asymmetry of Sec.~\ref{sec:asym} which may constitute an inhomogeneity spanning much of the observable Universe.

\subsection{Conclusion to Sec.~\ref{sec:foreground}: Lesson~III}

Alone through the meanwhile well-constrained ages and elemental abundances of stellar populations in nearby massive elliptical galaxies and the knowledge of their locations in co-moving volume, it is inescapable that the observed CMB is significantly contaminated by photons emitted during their formation. The correlation between the distribution of nearby elliptical galaxies with the CMB hemispherical asymmetry may imply that the Universe has an inhomogeneous structure that spans from here until $z\approx17$. The above makes it clear that the observed microwave background may have little to do with the surface of last scattering.

The results on the possible origin of the CMB as stemming from forming elliptical galaxies can be viewed in light of historic suggestions even though the present results are very distinct and rest on different physical arguments. Rees (1978 \cite{Rees1978}) discussed whether the observed CMB might not be related to pre-galactic stellar nucleosynthesis in the early Universe, and Burbidge \& Hoyle (1998 \cite{BurbidgeHoyle1998}) pointed out that the "energy released in the synthesis of cosmic $^4$He from hydrogen is almost exactly equal to the energy contained in the cosmic microwave background radiation" suggesting a stellar origin of both, the helium and the CMB. Vavrycuk (2018 \cite{Vavrycuk2018}) calculates the CMB to be a result of the thermal equilibrium at every $z$ between stellar light and dust observationally known to exist in inter-galactic space. 

To shed more light on the true nature and origin of the observed microwave background, the community needs to return to the raw-data collected by the WMAP and Planck missions to understand which signal from the young Universe is truly there. In doing so, different options for cosmological models and their structure formation histories need to be taken into account because each model leads to a different rate, strength and scale of structure formation.  It is important to be aware of the following problem: Assume there is no structure whatsoever in the true CMB produced at $z\approx 1100$, that the power-spectra are flat. By using the wrong cosmological model, the four corrections (Sec.~\ref{sec:unobservable}) will lead to apparent fluctuations and unphysical power spectra. 

The to-day claimed near-uniform temperature across the sky (within $10^{-5}$ level differences) as well as the currently-published CMB being a near-to-perfect black body with $T_{\rm obs}=2.73\,$K need to be revisited -- we need to understand what the data reduction that was applied with the Hot Big Bang SMoC in mind did to the detected microwave signal when the Planck-CMB data were being reduced. For each cosmological model this is an iterative procedure because each model leads to a different timing and strength of structures emerging, requiring the four model-specific corrections (Sec.~\ref{sec:unobservable}) to be repeatedly applied in order to iterate towards initial conditions that are compatible with the observed microwave background. If convergence cannot be achieved then the model may need to be discarded. 

It also needs to be understood how the spectral energy distributions of the forming and evolving elliptical galaxies combine, how their stellar emission is thermalised by the dust that forms and is destroyed within them and how uniform across the sky this background is, whereby the large-scale ($>\,$few~Gpc) inhomogeneities and cosmological-model-specific expansion history need to be allowed for. An important question to address is if the forming elliptical galaxies might be directly detectable with, for example the ALMA facility, remembering to take into account that the redshift--time--distance relations will differ for the correct model of cosmology compared to the currently used SMoC relations. 

\section{Conclusions}
\label{sec:concs}

The above discussion is based on two parts: In the first part, which contains Sec.~\ref{sec:uniqueness} and~\ref{sec:signature}, the underlying assumption is that recombination occurred near $z\approx 1100$, that the observed CMB comes from that time, that the formation of galaxies at $z>10$ can be ignored and that elliptical galaxies did not form with the high metal abundances they are observed to have. Two lessons are learned from this part, namely~(I) that the Planck-CMB data can be fitted by very different cosmological models such that the SMoC is not unique as a solution to the CMB emission. And we learned~(II) that the presently published Planck-CMB data are only valid in connection with the SMoC, while the data reduction of the raw Planck-CMB data needs to be redone for any other cosmological model separately. This invalidates the use of the currently published Planck-CMB data as initial conditions for other cosmological models. Doing so may lead to results that are not physical. But we do not yet know how large the errors are, since the Planck-CMB data have only been deduced by assuming the SMoC is valid. Lesson~II thus overrides Lesson~I.

The third lesson (III) overrides the two prior ones by telling us that if we take into account that massive elliptical galaxies exist and thus had to have formed and that their stars had to enrich the forming galaxies to the high and mostly super-Solar abundances they are observed to have, then the physical nature of the CMB cannot be that stemming from the recombination in a Hot Big Bang model. Even with the most conservative assumption the forming ellipticals contribute at least about 1.4~per cent to the observed CMB photon energy density. Since the elliptical galaxies are not smoothly distributed the $10^{-5}$-level temperature fluctuations in the CMB loose their physical meaning. The currently published Planck-CMB power spectra would thus appear to have little physical content. 

The deep implications of this deduction are ground breaking and require the community to reassess even fundamental ideas about the early Universe which were thought to have been established since many decades. The recent work related to the cosmological epoch of the formation of massive galaxies, and more fundamentally to the need to enrich the gas from which the stars in elliptcal galaxies formed to super-Solar metallicities, a process that released a photon energy density comparable to that observed in the CMB, thus opens the need to a very major new research effort that will require rethinking cosmological models from scratch. The ground-breaking discovery of the microwave background by Penzias\& Wilson (1965 \cite{PenziasWilson1965})\footnote{They estimated its temperature to be 3.5~K, but the CMB had been discovered using various tracers well before this officially-acknowledged discovery date. Hoyle, Burbidge \& Narlikar (2005 \cite{HBN2005} discuss evidence from the 1930s noting in particular the work of McKellar 1941 \cite{McKellar1941} based on absorption lines of the interstellar CN molecule. McKellar interpreted the required
population of the $J=1$ rotational level of the ground state of the molecule as being caused by radiative excitation from the $J=0$ level. The excitation radiation was taken to be from a black body, and the temperature required for it, in order to
explain the relative intensities of the observed lines, was 2.3~K. World War~II however was underway being one of the causes this work was not noted.} may thus constitute the detection of the first hyper-star bursts that lit up the young Universe with the faded remnants of this brief event being evident in the present-day massive elliptical galaxies. 

In brief, {\it this hitherto neglected extra-Galactic foreground necessitates a major revision of the physics of the early Universe and of cosmological theory in general}.

\begin{acknowledgments}
EG acknowledges the support of the National Natural Science Foundation of China (NSFC) under grants NOs. 1251101411, 12533003, 1257030642.
NS was supported by the Charles University Grant Agency (GAUK) with grant number~94224 and by the Ostrava Supercomputing Center with the Project OPEN-32-11 \#2885 as a PhD student and is currently employed under the FUND FD3026, grant number PG006935, cost center CC1714, at the University of Ottawa. PK and NS acknowledge the DAAD Eastern Europe Bonn-Prague exchange program at the University of Bonn and at Charles University for supporting the Bonn-Prague exchange visits.
\end{acknowledgments}
\bibliographystyle{JHEP}
\bibliography{refs_PKcorfu}{}
%

\end{document}